# A pre-metric generalization of the Lorentz transformation

D. H. Delphenich
Spring Valley, OH 45379

___

**Abstract:** The concept of an observer and their associated rest space is defined in a pre-metric (i.e., projective-geometric) context that relates to time+space decompositions of the tangent bundle to space-time. The transformation from one observer to another when the two are in a state of relative motion is then defined, and its relationship to the Lorentz transformation is discussed. The group of all linear transformations that preserve the observer quadric, which generalizes the proper-time hyperboloid in Minkowski space, is defined and the reductions to some of its subgroups are described, as well as its extension to the group that preserves the fundamental quadric, which generalizes the light cone.

**1. Introduction.** – Traditionally, special relativity [1] essentially begins with Minkowski space and the Lorentz group of transformations that preserve its scalar product. The concept of an observer is then introduced in the form of a Lorentzian frame – i.e., a linear frame that is orthonormal for the Minkowski scalar product. More precisely, one starts with a natural frame for a coordinate system on Minkowski space that is orthonormal for its scalar product. Concepts such as the *rest frame* and *rest space* for an observer are then typically introduced in a somewhat casual way that cries out for a more rigorous mathematical definition.

It will be shown in the following that one can start with a definition of an observer and their associated rest space in space-time that does not immediately require the introduction of the Minkowski scalar product, but still manages to allow one to define the concept of the time-line of the observer, as well as their rest space, and the concept of a rest frame emerges naturally, although it would be incorrect to say that it exists uniquely. One can then proceed to investigate the transformation between two observers that exist in a state of relative motion. Furthermore, the consistency of the methodology with the better-established methodology of special relativity becomes quite natural.

The key to understanding what is going on is to understand what distinguishes projective geometry from metric geometry. In Felix Klein's celebrated Erlanger Programm [2], geometries were defined by some fundamental construction or relationship and the group of transformations that preserved that construction or relationship. In the case of metric geometry, the fundamental construction is the metric on the set of points that one called "space," and the transformations then become isometries of that metric. By contrast, the fundamental relationship in projective geometry is that of "incidence." That is, whether a given geometric object, such as a point, line, plane, etc., was a subset of another geometric object or *vice versa*. It is important to see that this is distinct from mere intersection, since two lines can intersect at a point without necessarily being coincident. The transformations that preserve incidence are then projective transformations. Since one has essentially "symmetrized" the definition of incidence to mean that either one set of points is a subset of the other or the other way around, projective transformations will include duality transformations, which reverse the sense of set inclusion.

In Klein's picture of geometries, one can then classify geometries by their groups of transformations. Indeed, in that scheme, the group of projective transformations of a projective space then become the ultimate geometric group, while the groups of transformations for all of the other geometries became subgroups of it. For instance, one could go from projective geometry to affine, elliptic, or hyperbolic geometry by introducing an appropriate "absolute quadric" on the points at infinity. That probably explains the somewhat-apocryphal observation of Sir Arthur Cayley that "all geometry is projective geometry." Indeed, Klein also discussed the projective-geometric setting for the Lorentz group [**3**].

Some of the key concepts that will be emphasized in this study will be time+space decompositions of a vector space *V* and frames that are adapted to them. The main extension from Minkowski space will be from the proper-time hyperboloids and the light cone to some corresponding seven-dimensional quadric hypersurfaces in the Cartesian product of *V* with its dual space.

The general plan of what follows is to begin in section **2** by reviewing some basic aspects of Minkowski space and the Lorentz group, but in a slightly more "basis-free" sort of way. Section **3** will then define the concept of an "observer" in terms of time+space splittings of a vector space, which will also allow one to define the time-line and rest space of an observer, as well as the concept of rest frames. Section **4** will then get into the core material that is concerned with transformations of observers in the more general sense and the conditions under which they can reduce to Lorentz boosts. In the final section, the group of transformations that preserve the fundamental quadric that is defined by all observers will be discussed.

Although the scope of the following discussion is mostly restricted to four-dimensional vector spaces as a generalization of Minkowski space, nevertheless, its application to more general space-time manifolds is immediate. One simply assumes that the constructions are being made in the tangent and cotangent spaces to the more general manifold, in the same way that Minkowski space gives way to Lorentzian manifolds.

**2. Minkowski space.** – For the sake of completeness, we shall review some of the basic definitions, notations, and conventions that apply to the established geometry of Minkowski space in a way that will lead naturally into the more general constructions. For a review of the approach that will be taken to linear algebra, one can confer, e.g., Hoffman and Kunze [**4**].

*a. Basic definitions.* – For us, Minkowski space is a pair (*V*, $\eta$) that consists of a four-dimensional real vector space *V* and a symmetric, non-degenerate, bilinear form $\eta$ that then takes any pair of vector (**v**, **w**) in *V* to a real number $\eta$ (**v**, **w**), so one will also have:

(2.1) Symmetry: $\eta(\mathbf{v}, \mathbf{w}) = \eta(\mathbf{w}, \mathbf{v})$,

(2.2) Non-degeneracy: $\eta(\mathbf{v}, \mathbf{w}) = 0$ for all **w** iff **v** = 0,

(2.3) Bilinearity: $\eta(\alpha \mathbf{u} + \beta \mathbf{v}, \mathbf{w}) = \alpha\, \eta(\mathbf{u}, \mathbf{w}) + \beta\, \eta(\mathbf{v}, \mathbf{w})$.

(Since we have already assumed symmetry, it becomes redundant to specify the linearity in the second position – i.e., **w**.)

So far, we have defined only a scalar product on *V*, and sometimes we shall use the alternate notation:

(2.4) $$< \mathbf{v}, \mathbf{w} > \equiv \eta(\mathbf{v}, \mathbf{w}).$$

That makes *V* an *orthogonal space*, by definition.

A linear frame on *V* is a set $\{\mathbf{e}_\mu, \mu = 0, 1, 2, 3\}$ of four linearly-independent vectors in *V*. Hence, any vector $\mathbf{v} \in V$ can be expressed uniquely as a linear combination of those vectors:

(2.5) $$\mathbf{v} = v^\mu \mathbf{e}_\mu.$$

The real numbers $v^\mu$ are called the *components* of **v** relative to that choice of frame.

Any choice of linear frame on *V* will define a linear isomorphism of $\mathbb{R}^4$ with *V* that takes the *canonical frame vectors* $\delta_\mu = \{(1, 0, 0, 0), \ldots (0, 0, 0, 1)\}$ for $\mathbb{R}^4$ to the frame $\mathbf{e}_\mu$ for *V*. Hence, any vector $(v^0, v^1, v^2, v^3)$ in $\mathbb{R}^4$ will go to the vector $\mathbf{v} = v^\mu \mathbf{e}_\mu$ in *V*. Of course, a different choice of frame will take the $v^\mu$ to a different vector, so the linear isomorphism is by no means unique.

A linear frame $\mathbf{e}_\mu$ is called *orthonormal* or *Lorenzian* if one has:

(2.6) $$<\mathbf{e}_\mu, \mathbf{e}_\nu> = \eta(\mathbf{e}_\mu, \mathbf{e}_\nu) = \eta_{\mu\nu} \equiv \text{diag}[+1, -1, -1, -1].$$

It is the choice of the matrix in the final right-hand side of (2.6) that specifies Minkowski space, and one calls that matrix the *signature type* of the scalar product. In particular, one says that $\eta$ has a *normal hyperbolic* signature type. (One should be advised that the opposite sign convention is also used, as well as the "imaginary time" convention that makes the matrix take the Euclidian form of an identity matrix.)

Any linear frame $\{\mathbf{f}_\mu, \mu = 0, 1, 2, 3\}$ can be deformed into an orthonormal frame $\mathbf{e}_\mu$ by an invertible linear transformation due to the fact that each vector $\mathbf{f}_\mu$ can be expressed in terms of its components with respect to the frame $\mathbf{e}_\mu$:

(2.7) $$\mathbf{f}_\mu = \mathbf{e}_\nu f^\nu_\mu.$$

However, the customary Gram-Schmidt algorithm for starting with $\mathbf{f}_\mu$ and defining a unique $\mathbf{e}_\mu$ breaks down whenever one encounters light-like vectors, which cannot be normalized.

From bilinearity, one can express the scalar product of two vectors **v** and **w** in the component form when they are expressed in an orthonormal frame:

(2.8) $$<\mathbf{v}, \mathbf{w}> = \eta_{\mu\nu} v^\mu w^\nu.$$

For any frame that is not orthonormal, the component matrix for $\eta$ will not generally take the diagonal form in (2.6), so we will denote it by $g_{\mu\nu}$, and relative to that frame, we will have:

$$(2.9) \qquad <\mathbf{v}, \mathbf{w}> = g_{\mu\nu} v^\mu w^\nu.$$

The dual space to $V$ is the four-dimensional real vector space $V^*$ whose elements are linear functionals on $V$ that is, if $\alpha \in V^*$ and $\lambda \mathbf{v} + \mu \mathbf{w}$ is any linear combination of vectors in $V$ then $\alpha(\lambda \mathbf{v} + \mu \mathbf{w})$ will be a real number, and one must have:

$$(2.10) \qquad \alpha(\lambda \mathbf{v} + \mu \mathbf{w}) = \lambda \alpha(\mathbf{v}) + \mu \alpha(\mathbf{w}).$$

A vector in $V^*$ will also be referred to as a *covector*.

A *coframe* on $V^*$ is then a set $\{\theta^\mu, \mu = 0, 1, 2, 3\}$ of four linearly-independent covectors in $V^*$, so any covector $\alpha \in V^*$ can be expressed uniquely as a linear combination of the coframe members:

$$(2.11) \qquad \alpha = \alpha_\mu \theta^\mu.$$

The real numbers $\alpha_\mu$ are then the *components* of $\alpha$ relative to that choice of coframe.

A choice of coframe on $V^*$ will define a linear isomorphism of $V^*$ with $\mathbb{R}^4$ that takes any covector $\alpha$ to the row vector $(\alpha_0, \alpha_1, \alpha_2, \alpha_3)$ of its components. Similarly, it will define an isomorphism of $V$ with $\mathbb{R}^4$ that takes and vector $\mathbf{v} \in V$ to $\theta^\mu(\mathbf{v}) = v^\mu$ in $\mathbb{R}^4$.

Any linear frame $\mathbf{e}_\mu$ on $V$ gives rise to a unique coframe $\theta^\mu$ on $V^*$ that is called its *reciprocal coframe* and is defined by the rule:

$$(2.12) \qquad \theta^\mu(\mathbf{e}_\nu) = \delta^\mu_\nu,$$

in which $\delta^\mu_\nu$ is the usual Kronecker delta symbol, which equals 1 when $\mu = \nu$ and 0 otherwise; i.e., as a matrix, it is the identity matrix for $\mathbb{R}^4$. (2.12) also says that the linear isomorphism of $V$ with $\mathbb{R}^4$ that is defined the frame $\theta^\mu$ is the inverse of the linear isomorphism of $\mathbb{R}^4$ with $V$ that is defined by $\mathbf{e}_\mu$. (The fact that the reciprocal coframe is unique comes from the fact that any linear map between vector spaces is defined uniquely by its values on a chosen frame in the source vector space.)

Hence, a choice of $\mathbf{e}_\mu$ will also define a linear isomorphism of $V$ with $V^*$ that takes the frame $\mathbf{e}_\mu$ to the coframe $\theta^\mu$ and all other vector $\mathbf{X}$ in $V$, whose components with respect to $\mathbf{e}_\mu$ are $X^\mu$ will go to the covector $\mathbf{X}^\mathbf{T} = X_\mu \theta^\mu$, with $X_\mu = X^\mu$. In effect, all that one has done is to transpose the column vector whose components $X^\mu$ are to the row vector with the same components; hence, the notation T for the transpose of a matrix.

However, a different choice of frame and reciprocal coframe will define a different linear isomorphism of *V* its dual. That is because the change of $\mathbf{e}_\mu$ to $\mathbf{f}_\mu$ will have to be accompanied by the *inverse* transformation of $\theta^\mu$ in order for the resulting coframe to still be reciprocal to $\mathbf{f}_\mu$:

(2.13) $$\mathbf{f}_\mu = \mathbf{e}_\nu \, f_\mu^\nu, \qquad \vartheta^\mu = \tilde{f}_\nu^\mu \, \theta^\nu,$$

in which the tilde signifies the matrix inverse. That means that if $X^\mu$ are the components of $\mathbf{X}$ with respect to the new frame $\mathbf{f}_\mu$ then even though the components of $\mathbf{X}^T$ with respect to $\vartheta^\mu$ will still be $X_\mu = X^\mu$, nonetheless the components of $\mathbf{X}$ and $\mathbf{X}^T$ with respect to $\mathbf{e}_\mu$ and $\theta^\mu$, resp., will be $f_\nu^\mu X^\nu$ and $X_\nu \, \tilde{f}_\mu^\nu$, resp., which are not mere transposes of each other.

Since Minkowski space also has a scalar product defined on it, one can define another linear isomorphism of *V* with $V^*$ that is unique and independent of any choice of frame. One simply takes every vector $\mathbf{v}$ in *V* to the linear functional $\mathbf{v}^*$ that will give:

(2.14) $$\mathbf{v}^*(\mathbf{w}) = \eta(\mathbf{v}, \mathbf{w})$$

when it is evaluated on any vector $\mathbf{w}$.

When one chooses an orthonormal frame for *V* and its reciprocal coframe for $V^*$, one can represent $\mathbf{v}$ in the form (2.5) and $\mathbf{v}^*$ in the form:

(2.15) $$\mathbf{v}^* = v_\mu^* \, \theta^\mu, \quad \text{with} \quad v_\mu^* = \eta_{\mu\nu} v^\nu.$$

Hence, the isomorphism that $\eta$ defines is just what is commonly called "lowering the index" in the conventional literature of relativity. One can also say that the matrix of the scalar product is the matrix of the isomorphism. The inverse isomorphism of $V^*$ with *V* then takes the form of raising the index, and the matrix of that inverse is denoted by $\eta^{\mu\nu}$, which has the same components as $\eta_{\mu\nu}$ as a matrix. The matrix $\eta^{\mu\nu}$ then defines a scalar product on $V^*$ that is easiest to describe in a Lorentzian frame:

(2.16) $$\eta(\alpha, \beta) = \eta^{\mu\nu} \alpha_\mu \beta_\nu.$$

However, when one is dealing with Minkowski space, the fact that the signature type is not definite (i.e., all of one sign) implies that the coframe $\mathbf{e}^*_\mu$ that is metric-dual to the orthonormal frame $\mathbf{e}_\mu$ will differ from its reciprocal frame $\theta^\mu$ by three sign changes due to the fact that:

(2.17) $$\mathbf{e}^*_\mu(\mathbf{e}_\nu) = <\mathbf{e}_\mu, \mathbf{e}_\nu> = \eta_{\mu\nu}, \qquad \text{but} \qquad \theta^\mu(\mathbf{e}_\nu) = \delta_\nu^\mu.$$

Thus, if $\mathbf{v}$ is a vector in V then the components ($v_\mu = v^\mu$) of $\mathbf{v}^T$ relative to some frame $\mathbf{e}_\mu$ and its reciprocal coframe $\theta^\mu$ will also differ by three sign changes from those ($v_\mu = \eta_{\mu\nu} v^\mu$) of $\mathbf{v}^*$.

When $V$ has a scalar product defined on it, one can define a corresponding scalar product on $V^*$ by using the duality map $* : V^* \to V$ that comes from the scalar product on $V$:

(2.18) $$<\alpha, \beta> \equiv <\alpha^*, \beta^*>.$$

Hence, one can define orthonormality for a coframe $\theta^\mu$ on $V^*$ analogously:

(2.19) $$<\theta^\mu, \theta^\nu> = \eta^{\mu\nu} = \text{diag}\,[+1, -1, -1, -1].$$

One notes that the component matrix $\eta^{\mu\nu}$ is the inverse to the component matrix $\eta_{\mu\nu}$.

*b. Quadric hypersurfaces defined in Minkowski space.* – The scalar product $\eta$ allows one to define a corresponding quadratic form $Q\,[\mathbf{v}]$ on Minkowski space by way of:

(2.20) $$Q\,[\mathbf{v}] = <\mathbf{v}, \mathbf{v}> = \eta_{\mu\nu}\, v^\mu v^\nu = (v^0)^2 - (v^1)^2 - (v^2)^2 - (v^3)^2.$$

That defines three types of vectors in Minkowski space:

*a)*     Time-like:    $Q\,[\mathbf{v}] > 0$.

*b)*     Light-like:   $Q\,[\mathbf{v}] = 0$.

*c)*     Space-like:   $Q\,[\mathbf{v}] < 0$.

The set of all $\mathbf{v}$ such that $Q\,[\mathbf{v}]$ is the same constant then defines a quadric hypersurface in $V$. Depending upon whether that constant is positive, zero, or negative, it will be referred to as *time-like*, *light-like*, or *space-like*, respectively.

In particular, when:

(2.21) $$Q\,[\mathbf{v}] = c_0^2,$$

where $c_0$ is the speed of light *in vacuo*, the vectors $\mathbf{v}$ are all time-like and the quadric is a hyperboloid of two sheets that one calls the *proper-time hyperboloid*. Its two sheets are then called the *future* and *past* sheets, although the choice of which to call future or past is arbitrary and is referred to as a *time orientation*.

When:

(2.22) $$Q\,[\mathbf{v}] = 0,$$

the quadric will be a spherical cone through the origin that one calls the *time cone* of Minkowski space. Relative to an orthonormal frame, it will then take the forms:

(2.23) $\qquad 0 = (v^0)^2 - (v^1)^2 - (v^2)^2 - (v^3)^2 \qquad$ or $\qquad (v^0)^2 = (v^1)^2 + (v^2)^2 + (v^3)^2.$

In the latter form, one sees that it can be regarded as a one-parameter family of concentric 2-spheres of radius $v^0$. The light cone also has a future and a past sheet (when one deletes the origin), but the choice of which is which is also arbitrary. However, the future sheet of the proper-time hyperboloid must be interior to the future light cone, for consistency.

*b. Time+space splittings of Minkowski space.* – Whenever one has a time-like vector **u** in Minkowski space, one can generate a line through the origin [**u**] by way of all its scalar multiples. We shall call that a *time-line*. Because Minkowski space has a scalar product, one can then define a complementary hyperplane Σ that is orthogonal to [**u**], which consists of all vectors **v** ∈ V that are orthogonal to **u**; hence:

(2.24) $\qquad\qquad < \mathbf{u}, \mathbf{v} > = 0 .$

We shall call that orthogonal hyperplane the *rest space* of **u**. The vectors of Σ all become space-like with respect to the Minkowski scalar product, but that will be easier to see once we have introduced frames that are adapted to a time+space splitting, which we shall now define.

Since the line [**u**] and the hyperplane Σ intersect only at the origin and their dimensions add up to four, they collectively define a direct-sum splitting of $V = [\mathbf{u}] \oplus \Sigma$ that we shall call a *time+space* splitting of *V relative to* **u**. A different choice of **u** that is not collinear with the first one would then product a different direct-sum splitting.

A linear frame $\mathbf{e}_\mu$ on *V* is called *adapted* to the direct-sum [**u**] ⊕ Σ when one of its members (we shall always use $\mathbf{e}_0$) generates the line [**u**] by all of its scalar multiples, and the other three {$\mathbf{e}_i$, $i$ = 1, 2, 3} span the hyperplane Σ. When a frame that is adapted to **u** is also Lorentzian, one will also call it a *rest frame* for **u**. There will then be as many rest frames for **u** as there are linear frames on Σ, so we shall not give in to the popular temptation to refer to *the* rest frame of **u**.

When the scalar product on Minkowski space is restricted to pairs of vectors in a spatial hyperplane Σ, it will define a scalar product on Σ. In fact, when $\mathbf{v} = \mathbf{v}_t + \mathbf{v}_s$, $\mathbf{w} = \mathbf{w}_t + \mathbf{w}_s$, since $\mathbf{v}_t$ is orthogonal to $\mathbf{w}_s$ and $\mathbf{w}_t$ is orthogonal to $\mathbf{v}_s$, one can express the scalar product of **v** and **w** as:

(2.25) $\qquad\qquad < \mathbf{v}, \mathbf{w} > = < \mathbf{v}_t, \mathbf{w}_t > + < \mathbf{v}_s, \mathbf{w}_s > = v_t w_t c_0^2 + < \mathbf{v}_s, \mathbf{w}_s > .$

When **v** and **w** are expressed relative to an adapted Lorentzian frame $\mathbf{e}_\mu$, that will make:

(2.26) $\qquad\qquad < \mathbf{v}, \mathbf{w} > = v^0 w^0 c_0^2 - \delta_{ij} v^i w^j .$

Thus, one sees that the scalar product that is induced on Σ is minus the Euclidian scalar product. Hence, the vectors in Σ will all be space-like with respect to $< \cdot ., \cdot >$.

When *V* has been given a time+space splitting, any vector **X** ∈ *V* can be expressed uniquely as a sum $\mathbf{X}_t + \mathbf{X}_s$ where $\mathbf{X}_t = X_t \mathbf{u}$ for some unique real number $X_t$, and $\mathbf{X}_s$ is a unique vector that belongs to $\Sigma$. $\mathbf{X}_t$ is then called the *temporal part* of $\mathbf{X}$, while $\mathbf{X}_s$ is its *spatial part*, and $X_t$ is its *temporal component relative to* $\mathbf{u}$.

One can use the scalar product $\eta$ to define $X_t$, and if one defines $<\mathbf{u}, \mathbf{u}> = c_0^2$ then since:

$$<\mathbf{X}, \mathbf{u}> = <\mathbf{X}_t, \mathbf{u}> + <\mathbf{X}_s, \mathbf{u}> = X_t <\mathbf{u}, \mathbf{u}> = X_t \, c_0^2,$$

one will have:

(2.27) $$X_t = \frac{1}{c_0^2} <\mathbf{X}, \mathbf{u}>.$$

Hence, that will make:

(2.28) $$\mathbf{X}_s = \mathbf{X} - \mathbf{X}_t = \mathbf{X} - \frac{1}{c_0^2} <\mathbf{X}, \mathbf{u}> \mathbf{u}.$$

If we replace $<\mathbf{X}, \mathbf{u}>$ with $\mathbf{u}^*(\mathbf{X})$ then we can express $\mathbf{X}_t$ and $\mathbf{X}_s$ this in the form:

(2.29) $$\mathbf{X}_t = \frac{1}{c_0^2}(\mathbf{u}^* \otimes \mathbf{u})(\mathbf{X}), \qquad \mathbf{X}_s = (I - \frac{1}{c_0^2}\mathbf{u}^* \otimes \mathbf{u})(\mathbf{X}),$$

and that, in turn, will allow us to define temporal and spatial projection operators:

(2.30) $$P_t = \frac{1}{c_0^2}\mathbf{u}^* \otimes \mathbf{u}, \qquad P_s = I - \frac{1}{c_0^2}\mathbf{u}^* \otimes \mathbf{u},$$

which will then make:

(2.31) $$P_t(\mathbf{X}) = \mathbf{X}_t, \quad P_s(\mathbf{X}) = \mathbf{X}_s.$$

$P_t$ and $P_s$ have the characteristic properties of projection operators, namely:

(2.32) $$P_t \cdot P_t = P_t, \quad P_s \cdot P_s = P_s, \quad P_t \cdot P_s = P_s \cdot P_t = 0, \quad I = P_t + P_s,$$

in which $I$ is, of course, the identity operator on $V$.

Relative to a linear frame on $V$, the matrices of the projection operators will look like:

(2.33) $$[P_t]_\nu^\mu = \frac{1}{c_0^2} u_\nu u^\mu, \qquad [P_s]_\nu^\mu = \delta_\nu^\mu - \frac{1}{c_0^2} u_\nu u^\mu,$$

in which $u_\nu = g_{\nu\kappa} u^\kappa$.

Since $\eta$ will also define a scalar product on $V^*$, the dual covector $\mathbf{u}^*$ will also define a direct-sum splitting $V^* = [\mathbf{u}^*] \oplus \Sigma^*$. Analogous projection operators on $V^*$ will be defined in that way.

When a vector space *V* is given a direct-sum decomposition into $[\mathbf{u}] \oplus \Sigma$ that is analogous to a time+space splitting of Minkowski space, and $\Sigma$ is given a scalar product $<\cdot, \cdot>_\Sigma$, one can extend it to a scalar product on *V* by using that fact that any two vectors in *V* can be represented in the form $\alpha\mathbf{u} + \mathbf{v}$, $\beta\mathbf{u} + \mathbf{w}$, where $\mathbf{v}$ and $\mathbf{w}$ belong to $\Sigma$, so from the bilinearity of any scalar product and the fact that both $\mathbf{v}$ and $\mathbf{w}$ are assumed to be orthogonal to $\mathbf{u}$, one must have:

(2.34) $\qquad <\alpha\mathbf{u} + \mathbf{v}, \beta\mathbf{u} + \mathbf{w}> = \alpha\beta <\mathbf{u}, \mathbf{u}> + <\mathbf{v}, \mathbf{w}>.$

Thus, when one lets $<\mathbf{v}, \mathbf{w}> = <\mathbf{v}, \mathbf{w}>_\Sigma$, the only thing that is left to be defined is $<\mathbf{u}, \mathbf{u}>$, which can then be any non-zero real number. If $<\mathbf{v}, \mathbf{w}>_\Sigma$ is the (negative) Euclidian scalar product on $\Sigma$ then choosing $<\mathbf{u}, \mathbf{u}> = c_0^2$ would give the usual Minkowski scalar product on *V*.

*c. Lorentz transformations.* – When *V* is Minkowski space, one can define a *Lorentz transformation* to be a linear map $L : V \to V$ that preserves the scalar product. Hence, for any pair of vectors $\mathbf{v}, \mathbf{w}$ in *V* one must have:

(2.35) $\qquad <L(\mathbf{v}), L(\mathbf{w})> = <\mathbf{v}, \mathbf{w}>.$

In matrix form, this is:

(2.36) $\qquad \mathbf{v}^T L^T \eta L \mathbf{w} = \mathbf{v}^T \eta \mathbf{w},$

so if that is true for every choice of $\mathbf{v}$ and $\mathbf{w}$ then one must have:

(2.37) $\qquad L^T \eta L = \eta \quad \text{or} \quad \eta L^T \eta L = I,$

since the matrix of $\eta$ is its own inverse. That also means that the inverse of *L* will be:

(2.38) $\qquad L^{-1} = \eta L^T \eta.$

One might compare this to the Euclidian case, in which $\eta$ gets replaced with the identity matrix, so the inverse of an orthogonal transformation becomes its transpose.

The set of all Lorentz transformations, given the binary operation of composition of operators, becomes a group that is commonly called the *Lorentz group*. Because such transformations preserve scalar products, they will also preserve the associated quadratic form $Q[\cdot]$, and as a result, they will take time-like vectors to time-like vectors, light-like vectors to light-like ones, and space-like vectors to space-like ones; i.e., they will preserve the quadric hypersurfaces that are defined by *Q*. They will also take Lorentzian frames to other such things.

When *V* is given a time+space splitting relative to a time-like $\mathbf{u}$, the restriction of any Lorentz transformation to $\Sigma$ will be an orthogonal transformation of the Euclidian space that is defined on it. Thus, the Lorentz group includes all (proper and improper) spatial rotations.

However, it also includes the transformations that are sometimes called Lorentz transformations, in their own right, but they are also called *boosts*, which is the term that we shall employ. They define the essential difference between Galilean relativity and special relativity, as Einstein envisioned it, since they relate to the transformation between two Lorentzian frames that differ by being a state of relative uniform motion with a relative velocity of **v**. The prototype of all such transformations can be defined on the Minkowski plane; i.e., the plane of **u** and **v**, where **u** is time-like and **v** is spatial. For our purposes, it will be sufficient to represent it in the form:

$$(2.39) \qquad \mathbf{e}'_0 = \gamma \left( \mathbf{e}_0 - \frac{v}{c_0} \mathbf{e}_1 \right), \qquad \mathbf{e}'_1 = \gamma \left( -\frac{v}{c_0} \mathbf{e}_0 + \mathbf{e}_1 \right),$$

in which the Lorentzian frame $\{\mathbf{e}_0, \mathbf{e}_1\}$ is defined by the unit vectors in the directions of **u** and **v**, respectively, while:

$$(2.40) \qquad \gamma = \left( 1 - \frac{v^2}{c_0^2} \right)^{-1/2}, \qquad (v^2 \equiv - <\mathbf{v}, \mathbf{v}>)$$

is the ubiquitous *Fitzgerald-Lorentz* factor, which accounts for such experimentally-observed phenomena as time dilation, length contraction, and the increase in relative mass.

In a sense, the boost in the direction of **v** that is defined by (2.39) is the "canonical" form for a boost, since any other boost in a different direction **v**′ with the same value of $v^2$ can be brought into that form by a suitable rotation of the spatial axes. Hence, there are as many boosts as spatial rotations, which define a three-dimensional Lie group, and that makes the Lorentz group six-dimensional. However, the boosts by themselves do not define a group, except when one considers all boosts in the same direction (i.e., all $0 \leq v < c_0$). Once again, that is because the composition of two boosts in different directions can be factored into the product of a boost and a rotation. (That is the basis for "Thomas precession.")

**3. Space-time observers and time+space splittings.** – We shall now introduce the concept of a space-time observer in a manner that does not require the introduction of a Minkowski scalar product. After that, we shall show how one might transform between observers and then show how it would relate to the Lorentz transformations. There is an appreciable volume of literature by now on the subject of observers, time+space splittings, transverse geometry, and the like, and many references can be found in the author's paper [**5**].

*a. Basic definitions.* – Space-time, for us, shall be a four-dimensional real vector space $V$, and its dual vector space shall be denoted by $V^*$. Since the elements of $V^*$ are linear functionals on vectors in $V$, there is a natural – i.e., *canonical* – bilinear pairing of $V$ and its dual that produces a real scalar by evaluating a functional on a vector:

$$(3.1) \qquad V^* \times V \to \mathbb{R}, \quad (\alpha, \mathbf{X}) \mapsto \alpha(\mathbf{X}).$$

If one introduces a linear frame $\{\mathbf{e}_\mu, \mu = 0, 1, 2, 3\}$ on $V$ and its reciprocal coframe $\{\theta^\mu, \mu = 0, 1, 2, 3\}$ on $V^*$ then one can then express the vector $\mathbf{X}$ as a linear combination $X^\mu \mathbf{e}_\mu$ and the covector $\alpha$ in the form $\alpha_\mu \theta^\mu$, and one will then have:

(3.2) $$\alpha(\mathbf{X}) = \alpha_\mu X^\mu.$$

Any non-zero vector $\mathbf{X}$ in $V$ defines a line $[\mathbf{X}]$ through the origin by way of all of its scalar multiples. It also defines a hyperplane $\Sigma^*$ in $V^*$ by way of all linear functionals that annihilate it; i.e.:

(3.3) $$\Sigma^* = \{\text{all } \alpha \text{ such that } \alpha(\mathbf{X}) = 0\}.$$

Dually, any non-zero covector $\alpha$ will generate a line $[\alpha]$ through the origin of $V^*$ and a hyperplane $\Sigma$ in $V$ that consists of all vectors $\mathbf{v}$ that are annihilated by $\alpha$.

The key to understanding the present discussion is to see that the algebraic relationship $\alpha(\mathbf{X}) = 0$ expresses the geometric relationship of incidence in both cases of $\Sigma$ and its dual $\Sigma^*$. That is, when $\alpha(\mathbf{X}) = 0$, the vector $\mathbf{X}$ will be incident on the hyperplane that is defined by $\alpha$. Hence, one is dealing with a fundamentally projective-geometric concept.

In the event that the vector space is Minkowski space, so it also has a scalar product $<\cdot,\cdot>$ defined on it, the natural linear isomorphism of $V$ with $V^*$ that takes any vector $\mathbf{X}$ in $V$ to a covector $\mathbf{X}^*$ in $V^*$ will make:

(3.4) $$\mathbf{X}^*(\mathbf{Y}) = <\mathbf{X}, \mathbf{Y}> = \eta_{\mu\nu} X^\mu Y^\nu = X_\mu Y^\mu$$

for any vector $\mathbf{Y}$ in $V$. Dually, any covector $\alpha$ in $V^*$ can be associated with a vector $\alpha^* = (\eta^{\mu\nu} \alpha_\nu) \mathbf{e}_\mu$ that makes the bilinear pairing of $V$ and $V^*$ take the form:

(3.5) $$(\alpha, \mathbf{X}) = <\alpha^*, \mathbf{X}> = \eta_{\mu\nu} \alpha^\mu X^\nu.$$

This has the effect of saying that incidence in Minkowski space is related to orthogonality. That is, $\mathbf{X}$ is incident on the hyperplane $\Sigma$ that is annihilated by $\alpha$ iff $\mathbf{X}$ is orthogonal to the vector $\alpha^*$. The hyperplane $\Sigma$ then becomes the orthogonal complement to the line $[\alpha^*]$. One sees that the essential generalization from orthogonal spaces to projective geometry amounts to replacing the scalar product of vectors with the canonical bilinear pairing of vectors and covectors. The expansion of scope is due to the fact that not all linear isomorphisms of $V$ with $V^*$ can be represented in the form of metric isomorphisms, and that amounts to saying that the bilinear form on $V$ that is defined by a linear isomorphism $\iota: V \to V^*$, namely:

(3.6) $$<\mathbf{X}, \mathbf{Y}> = \iota(\mathbf{X})(\mathbf{Y}) = \iota_{\mu\nu} X^\mu Y^\nu,$$

does not have to be symmetric in **X** and **Y** ; i.e., the component matrix $\iota_{\mu\nu}$ does not have to be symmetric in its indices. In projective geometry, such an isomorphism of $V$ with $V^*$ (or rather, its projection onto the projective spaces $PV$ and $PV^*$) is called a *correlation*.

Our definition of a *space-time observer* will be simply a pair $(u, \mathbf{u})$ that consists of a vector $\mathbf{u}$ in $V$ and a covector $u$ in $V^*$ that are constrained by the demand that ([1]):

(3.7) $$u(\mathbf{u}) = u_\mu u^\mu = c_0^2$$

When one is dealing with Minkowski space, this will take the form of saying that **u** would be time-like and lie on the proper-time hyperboloid. However, since the relationship between **u** and $u$ is less specific than that of **u** and $\mathbf{u}^*$, one is no longer dealing with a three-dimensional quadric in $V$, but a seven-dimensional quadric in $V^* \times V$. In older literature [**6**], that quadric (or rather, its projection onto $PV^* \times PV$, where $PV$ and $PV^*$ are the projective spaces that are defined by the sets of lines through the origin in $V$ and $V^*$, defined what was called an *algebraic correspondence*. If one has a scalar product on $V$, however, one can see that this more-general quadric contains the proper-time quadric in the form of all pairs of the form $(\mathbf{u}^*, \mathbf{u})$. Because of the central role that it plays in the theory, we shall refer to the quadric that is defined by (3.7) as the *observer quadric*.

The algebraic correspondence between vectors and covectors that is defined by the observer quadric is hardly a one-to-one correspondence. Indeed, if a vector **u** satisfies (3.7) for some fixed $u$ then any vector **u**′ that differs from **u** by a vector $\mathbf{v} \in \Sigma$ will also satisfy it. Hence, a choice of $u$ will define only an affine hyperplane in $V$, but not a unique vector. Similarly, when one fixes **u**, $u$ will be defined only up to a covector in $\Sigma^*$, so a choice of **u** will be associated with an affine hyperplane in $V^*$. One can then regard these two affine hyperplanes as equivalence classes of vectors and covectors under the equivalences:

(3.8) $$\mathbf{u}' \cong \mathbf{u} \text{ iff } \mathbf{u}' - \mathbf{u} \in \Sigma, \quad u' \cong u \text{ iff } u' - u \in \Sigma^*.$$

One can also regard the affine hyperplane that corresponds to **u**′ as the translate $\mathbf{u} + \Sigma$ of $\Sigma$ by **u**, and similarly the affine hyperplane that corresponds to $u'$ is the translate $u + \Sigma^*$.

When one has an observer $(u, \mathbf{u})$, one can refer to the line $[\mathbf{u}]$ as the *time-line* of the observer and the hyperplane $\Sigma$ as the *rest space* of the observer. A *rest frame* then becomes a frame $\mathbf{e}_\mu$ that is adapted to those spaces in the sense that $\mathbf{e}_0$ generates the same line as **u**, and the set of three frame members $\{\mathbf{e}_i, i = 1, 2, 3\}$ spans the hyperplane $\Sigma$. Since a rest frame is not by any means unique at this point (due to the infinitude of linear frames on $\Sigma$), we shall not refer to "the" rest frame of an observer.

---

([1]) Although it might seem more concise to use 1 on the right-hand side of this relation, the introduction of the speed of light *in vacuo* $c_0$ is to make the consistency with the usual formulation of special relativity more straightforward.

*b. Time+space splittings of space-time.* – When one is given an observer $(u, \mathbf{u})$, with the present definition, the fact that $u(\mathbf{u})$ is not equal to zero says that the time-line $[\mathbf{u}]$ does not lie in the rest space $\Sigma$; i.e., they are transverse subspaces of $V$. Since their dimensions are complementary, one can then say that they define a *time+space* splitting of $V$ in the form of a direct-sum decomposition:

$$(3.9) \qquad V = [\mathbf{u}] \oplus \Sigma .$$

Hence, any vector $\mathbf{X}$ in $V$ can be expressed uniquely in the form:

$$(3.10) \qquad \mathbf{X} = \mathbf{X}_t + \mathbf{X}_s = X_t \mathbf{u} + \mathbf{X}_s ,$$

in which $\mathbf{X}_t$ belongs to $[\mathbf{u}]$ and $\mathbf{X}_s$ belongs to $\Sigma$. One again refers to $\mathbf{X}_t$ as the *temporal* part of $\mathbf{X}$ and $\mathbf{X}_s$ as its *spatial* part. The scalar $X_t$ is the *temporal component* of $\mathbf{X}$.

If a vector $\mathbf{X}$ is represented in time+space form, as in (3.10), then one will see that since $u$ annihilates all vectors in $\Sigma$ :

$$(3.11) \qquad u(\mathbf{X}) = u(\mathbf{X}_t) = u(X_t \mathbf{u}) = X_t c_0^2 ;$$

i.e.:

$$(3.12) \qquad X_t = \frac{1}{c_0^2} u(\mathbf{X}) .$$

Because of the uniqueness of that decomposition, one can once more define projection operators $P_t : V \to [\mathbf{u}]$ and $P_s : V \to \Sigma$, that take any $\mathbf{X} \in V$ to:

$$(3.13) \qquad P_t(\mathbf{X}) = \mathbf{X}_t , \quad P_s(\mathbf{X}) = \mathbf{X}_s ,$$

which once again satisfy the traditional properties of projection operators:

$$(3.14) \qquad P_t \cdot P_t = P_t , \quad P_s \cdot P_s = P_s , \quad P_t \cdot P_s = P_s \cdot P_t = 0, \quad I = P_t + P_s .$$

That allows one to represent the projection operators $P_t$ and $P_s$ in terms of $u$ and $\mathbf{u}$ as tensors of mixed type:

$$(3.15) \qquad P_t = \frac{1}{c_0^2} u \otimes \mathbf{u}, \qquad P_s = I - P_t = I - \frac{1}{c_0^2} u \otimes \mathbf{u} .$$

The component form of these expressions is still the same as in (2.33), but one will not generally have that $u = \mathbf{u}^*$ (i.e., $u_\mu = \eta_{\mu\nu} u^\nu$).

Dually, one has a time+space splitting of $V^*$ into a direct sum:

(3.16) $$V^* = [u] \oplus \Sigma^*$$

and a unique representation of any covector $\alpha$ in the form:

(3.17) $$\alpha = \alpha_t + \alpha_s = a_t\, u + \alpha_s,$$

with analogous terminology for the components. Similarly, one has projection operators that are defined by that unique decomposition.

When a covector-vector pair ($\xi$, **x**) has been expressed in time+space form relative to an observer ($u$, **u**) in the form:

(3.18) $$\mathbf{X} = \lambda\,(\mathbf{u} + \mathbf{v}), \qquad \xi = \tilde{\lambda}\,(u - v),$$

one will have:

(3.19) $$\xi(\mathbf{X}) = \lambda\tilde{\lambda}\,(c_0^2 - v^2),$$

since:

(3.20) $$u(\mathbf{u}) = c_0^2, \qquad u(\mathbf{v}) = v(\mathbf{u}) = 0, \qquad v(\mathbf{v}) \equiv v^2.$$

We can define another quadric by:

(3.21) $$\xi(\mathbf{X}) = 0,$$

which we shall call the *fundamental quadric*. Relative to the present choice of observer, as long as $\lambda$ and $\tilde{\lambda}$ are both non-vanishing (which would be equivalent to both **X** and $\xi$ being non-vanishing), it will be defined by all ($v$, **v**) such that:

(3.22) $$v(\mathbf{v}) = v^2 = c_0^2.$$

Hence, it can also be regarded as an affine quadric in $\Sigma \times \Sigma^*$.

**4. Transformation of observers.** – If one has two observers ($u$, **u**) and ($u'$, **u**$'$) then one will have two time+space decompositions of $V$ and $V^*$ accordingly. Hence, any vector **X** in $V$ can be written in two different ways:

(4.1) $$\mathbf{X} = \mathbf{X}_t + \mathbf{X}_s = \mathbf{X}'_t + \mathbf{X}'_s$$

depending upon which decomposition is used.

Similarly, any covector $\alpha$ in $V^*$ can be written in two different ways:

(4.2) $$\alpha = \alpha_t + \alpha_s = \alpha'_t + \alpha'_s.$$

In particular, one can express $\mathbf{u}'$ and $u'$ themselves in terms of $\mathbf{u}$ and $u$:

(4.3) $$\mathbf{u}' = \mathbf{u}'_t + \mathbf{u}'_s, \quad u' = u'_t + u'_s,$$

in which:

(4.4) $$\mathbf{u}'_t = \gamma \mathbf{u}, \quad u(\mathbf{u}'_s) = 0, \quad u'_t = \tilde{\gamma} u, \quad u'_s(\mathbf{u}) = 0.$$

Since both $(u, \mathbf{u})$ and $(u', \mathbf{u}')$ must lie on the observer quadric, we must have:

$$c_0^2 = u'(\mathbf{u}') = (\tilde{\gamma} u + u'_s)(\gamma \mathbf{u} + \mathbf{u}'_s) = \gamma \tilde{\gamma} c_0^2 + (u'_s)(\mathbf{u}'_s).$$

If we define the spatial vector $\mathbf{v}$ and the spatial covector $\upsilon$ to make:

(4.5) $$\mathbf{u}'_s = \gamma \mathbf{v}, \qquad u'_s = -\tilde{\gamma} \upsilon, \qquad \upsilon(\mathbf{v}) \equiv v^2$$

then we will see that we can solve for the product of the scalars:

(4.6) $$\gamma \tilde{\gamma} = \left(1 - \frac{v^2}{c_0^2}\right)^{-1}.$$

We immediately recognize the square of the Fitzgerald-Lorentz coefficient on the right-hand side of this. Hence, we have almost duplicated part of the Lorentz transformation between relativistic observer. However, since the left-hand side does not have to take the form of $\gamma^2$, in general, one can treat the case in which that is true as a special case. That is, $\gamma$ and $\tilde{\gamma}$ are independent, except for the constraint (4.6), just as $\mathbf{u}$ and $u$ are independent, except for the constraint (3.7). Moreover, $\gamma$ and $\tilde{\gamma}$ both become dependent upon $v$ in the process.

We can rewrite our decompositions of $\mathbf{u}'$ and $u'$ in terms of what we have established:

(4.7) $$\mathbf{u}' = \gamma (\mathbf{u} + \mathbf{v}), \qquad u' = \tilde{\gamma}(u - \upsilon).$$

That suggests that we can define any transformation of an observer to another observer by a pair $(\upsilon, \mathbf{v})$ such that $\mathbf{v}$ and $\upsilon$ are both spatial relative to the first observer $(u, \mathbf{u})$ and $\upsilon(\mathbf{v}) > 0$, along with a pair of non-zero scalars $\gamma$ and $\tilde{\gamma}$ that are coupled by the constraint in (4.6). Hence, since $\mathbf{v}$ and $\upsilon$ each have three components, while $\gamma$ and $\tilde{\gamma}$ add two more dimensions to the space of all transformations, but the condition (4.6) subtracts one dimension, we are left with seven. The condition that $\upsilon(\mathbf{v}) > 0$ does not reduce the dimension, since it is an inequality, not an equality, even when one adds the physically-motivated condition that $v < c_0$, which is also an inequality.

*a. Transformation of arbitrary vectors between observers*. – An arbitrary vector $\mathbf{X} \in \mathbb{R}^4$ can be expressed in two different ways relative to two different observers:

(4.8) $$\mathbf{X} = X_t \mathbf{u} + \mathbf{X}_s = X'_t \mathbf{u}' + \mathbf{X}'_s \, .$$

Our first problem is to express $X'_t$ and $\mathbf{X}'_s$ in terms of $X_t$ and $\mathbf{X}_s$ and the parameters of the basic transformation (4.7). We can first say that:

$$X'_t = \frac{1}{c_0^2} u'(\mathbf{X}) = \frac{\tilde{\gamma}}{c_0^2}(u - \upsilon)(X_t \mathbf{u} + \mathbf{X}_s)$$

or

(4.9) $$X'_t = \tilde{\gamma}[X_t - \frac{1}{c_0^2}\upsilon(\mathbf{X}_s)] \, .$$

We then have that:

$$\mathbf{X}'_s = \mathbf{X} - X'_t \mathbf{u}' = X_t \mathbf{u} + \mathbf{X}_s - \gamma\tilde{\gamma}[X_t - \frac{1}{c_0^2}\upsilon(\mathbf{X}_s)](\mathbf{u} + \mathbf{v})$$

$$= [(1 - \gamma\tilde{\gamma})\mathbf{u} - \gamma\tilde{\gamma}\,\mathbf{v}]X_t + \left[I + \frac{\gamma\tilde{\gamma}}{c_0^2}\upsilon \otimes (\mathbf{u} + \mathbf{v})\right](\mathbf{X}_s) \, .$$

Since:

$$1 - \gamma\tilde{\gamma} = 1 - \frac{c_0^2}{c_0^2 - v^2} = -\frac{v^2}{c_0^2 - v^2} = -\gamma\tilde{\gamma}\frac{v^2}{c_0^2} \, ,$$

we can say that:

(4.10) $$X'_t = \tilde{\gamma}[X_t - \frac{1}{c_0^2}\upsilon(\mathbf{X}_s)], \qquad \mathbf{X}'_s = -\gamma\tilde{\gamma}\left(\frac{v^2}{c_0^2}\mathbf{u} + \mathbf{v}\right)X_t + \left[I + \frac{\gamma\tilde{\gamma}}{c_0^2}\upsilon \otimes (\mathbf{u} + \mathbf{v})\right](\mathbf{X}_s) \, .$$

Hence, between these two equations, we can, in principle, express the transformation of temporal and spatial components of any vector under a change of observer. They also define an invertible linear transformation of those components.

However, the transformation that we defined does not take adapted frames to adapted frames. In particular, one notes that $\mathbf{u}$ ($X_t = 1$, $\mathbf{X}_s = 0$) goes to something whose temporal component is $\tilde{\gamma}$ and whose spatial component is $\mathbf{X}'_s = -\gamma\tilde{\gamma}\left(\frac{v^2}{c_0^2}\mathbf{u} + \mathbf{v}\right)$, which is not zero. Similarly, a purely spatial vector, such as $\mathbf{X}_s$ ($X_t = 0$), goes to something with a non-zero temporal part, namely, $-\frac{\tilde{\gamma}}{c_0^2}\upsilon(\mathbf{X}_s)$.

What we have is a way of expressing *the same* vector **X** in terms of two different frames, while what we want now is a way of transforming **X** into a *different* vector. In particular, we want to extend the transformation that takes **u** to $\gamma(\mathbf{u} + \mathbf{v})$. In terms of a frame change, that means that the vector $X^\mu \mathbf{e}_\mu$ will go to the vector $X'^\mu \mathbf{e}'_\mu$.

Let us now define two adapted frames $\mathbf{e}_\mu$ and $\mathbf{e}'_\mu$ on $\mathbb{R}^4$. By definition, $\mathbf{e}_0$ will be collinear with **u** and $\mathbf{e}'_0$ will be collinear with **u**′, while $\mathbf{e}_i$ will span $\Sigma$ and $\mathbf{e}'_i$ will span $\Sigma'$.

In anticipation of the eventual reduction to Lorentz transformations, we make the definitions and replacements:

(4.11) $$\mathbf{u} = c_0\, \mathbf{e}_0, \qquad \mathbf{u}' = c_0\, \mathbf{e}'_0, \qquad \mathbf{e}'_i \to \tilde{\gamma}\, \mathbf{e}'_i.$$

Since we already know how to transform **u**, we can now put it into the form:

(4.12) $$\mathbf{e}'_0 = \gamma\left(\mathbf{e}_0 + \frac{v^i}{c_0}\mathbf{e}_i\right).$$

The second of equations (4.10) can then be put into the form:

(4.13) $$X'^i \mathbf{e}'_i = -\gamma\tilde{\gamma}\left(\frac{v^2}{c_0}\mathbf{e}_0 + v^i \mathbf{e}_i\right) X^0 + \left[\mathbf{e}_i + \frac{\tilde{\gamma}}{c_0^2} v_i (c_0 \mathbf{e}_0 + v^j \mathbf{e}_j)\right] X^i.$$

This allows us to express any vector in $\Sigma'$ (for which $X^0 = 0$) in the form:

(4.14) $$X'^i \mathbf{e}'_i = \frac{\tilde{\gamma}}{c_0} v_i X^i \mathbf{e}_0 + \left(\delta_i^j + \frac{\tilde{\gamma}}{c_0^2} v_i v^j\right) X^i \mathbf{e}_j.$$

Hence, each basis vector $\tilde{\gamma}\,\mathbf{e}'_i$ ($X'^i = X^i = \delta^i_j$) can be expressed in the form:

(4.15) $$\tilde{\gamma}\,\mathbf{e}'_i = \frac{\tilde{\gamma}}{c_0} v_i\, \mathbf{e}_0 + \left(1 + \tilde{\gamma}\frac{v^2}{c_0^2}\right)\mathbf{e}_i.$$

The term in parentheses reduces to:

$$1 + \tilde{\gamma}\frac{v^2}{c_0^2} = \tilde{\gamma}\gamma,$$

so:

$$\tilde{\gamma}\,\mathbf{e}'_i = \gamma\tilde{\gamma}\left(\frac{v_i}{c_0}\mathbf{e}_0 + \mathbf{e}_i\right).$$

We then combine our formulas for transforming an adapted frame into an adapted frame into:

$$(4.16) \qquad \mathbf{e}'_0 = \gamma \left( \mathbf{e}_0 + \frac{v^i}{c_0} \mathbf{e}_i \right), \qquad \mathbf{e}'_i = \gamma \left( \frac{v_i}{c_0} \mathbf{e}_0 + \mathbf{e}_i \right).$$

*b. Reduction to Lorentz transformations.* – Recall the basic form for a Lorentz transformation (2.39):

$$(4.17) \qquad \mathbf{e}'_0 = \gamma (\mathbf{e}_0 - \frac{v}{c_0} \mathbf{e}_1), \qquad \mathbf{e}'_1 = \gamma (-\frac{v}{c_0} \mathbf{e}_0 + \mathbf{e}_1),$$

A comparison of these equations with (4.16) will show that one can make them take the same form by first setting $\gamma = \tilde{\gamma}$, so $\gamma$ will become the actual Fitzgerald-Lorentz factor, adapting the spatial frame $\mathbf{e}_i$ to the velocity vector $\mathbf{v} = v\,\mathbf{e}_1$, setting $v_i = v^i$, and changing the sign on $v$.

One sees that one difference between the transformation (4.16) and a Lorentz transformation is that there are two scalar multipliers $\gamma$ and $\tilde{\gamma}$ involved, rather than just the one, namely, $\gamma$. Another essential difference is the fact that in the case of Minkowski space a choice of $\mathbf{v}$ will imply a unique choice of $\upsilon$, while in the more general case of the fundamental quadric on $V^* \times V$, a choice of $\mathbf{v}$ will define only an affine hyperplane of covectors that could serve as $\upsilon$. Hence, whereas a choice of boost vector $\mathbf{v}$ in the Minkowski case will imply a single $\gamma$ and a unique Lorentz transformation, more generally, one must choose $\gamma$, $\tilde{\gamma}$, $\mathbf{v}$ and $\upsilon$ independently (as long as they are consistent with the fundamental constraints). That is why the more general transformation of observers behaves more like a pair of independent Lorentz transformations of $V$ and $V^*$, while defining a unique linear isomorphism of $V$ with $V^*$ will reduce that to just one.

*c. Dual transformations.* – When we express an arbitrary covector $\alpha \in \mathbb{R}^{4*}$ in the two forms:

$$(4.18) \qquad \alpha = \alpha_t\, u + \alpha_s = \alpha'_t\, u' + \alpha'_s, \qquad \text{with} \qquad u' = \tilde{\gamma}\,(u - v),$$

we can follow through the same sequence of calculations as in (4.9) to (4.16) with appropriate alterations and obtain analogous results. We shall simply summarize those results.

Equations (4.10) now take the form:

$$(4.19) \qquad \alpha'_t = \gamma [\alpha_t + \frac{1}{c_0^2} \alpha_s(\mathbf{v})], \qquad \alpha'_s = -\gamma \tilde{\gamma} \left( \frac{v^2}{c_0^2} u - v \right) \alpha_t + \left[ I - \frac{\gamma \tilde{\gamma}}{c_0^2} \mathbf{v} \otimes (u - v) \right](\alpha_s).$$

Upon introducing the coframes $\theta^\mu$ and $\theta'^\mu$, with $u = c_0\, \theta^0$ and $u' = c_0\, \theta'^0$, equation (4.15) becomes:

(4.20) $$\gamma \theta'^i = -\frac{\gamma \tilde{\gamma}}{c_0} v^i \theta^0 + \left(1 + \gamma \tilde{\gamma} \frac{v^2}{c_0^2}\right) \theta^i,$$

which makes the dual coframe transformations (4.16) now take the form:

(4.21) $$\theta'^0 = \tilde{\gamma}\left(\theta^0 - \frac{v_i}{c_0} \theta^i\right), \qquad \theta'^i = \tilde{\gamma}\left(\frac{v^i}{c_0} \theta^0 - \theta^i\right).$$

Analogous statements apply to their relationship to the Lorentz transformations.

**5. The group of linear transformations that preserve the canonical bilinear pairing.** – So far, we have considered only transformations of observers, which generalized the Lorentz boosts, but not the more general transformations that might generalize the Euclidian rotations in the rest space. Since an observer $(u, \mathbf{u})$ is an element of the vector space $V^* \times V$, we shall begin by considering the two types of linear transformations $V^* \times V \to V^* \times V$, $(\alpha, \mathbf{X}) \mapsto (\alpha', \mathbf{X}')$.

First, we have the transformations $(L_1, L_2)$ that belong to the direct product $GL(n) \times GL(n)$, so both $L_1$ and $L_2$ are invertible linear transformations of $V$. However, the action of $L_2$ on $V^*$ is by way of its transpose as a map $L_2 : V \to V$, namely, $L_2^T : V^* \to V^*$, $\alpha \mapsto L^T(\alpha)$, where:

(5.1) $$(L_2^T \alpha)(\mathbf{X}) = \alpha(L_2(\mathbf{X})).$$

When $\alpha$ is represented as row matrix and $L_2$ is an $n \times n$ matrix, the action in question will be simply the matrix multiplication of $\alpha$ times $L_2$.

Hence, the equations of transformation in this case will be the pair of transformations:

(5.2) $$\mathbf{X}' = L_1(\mathbf{X}), \qquad \alpha' = \alpha L_2.$$

A second type of linear transformation that takes $V^* \times V$ to itself is a duality transformation. The consist of pairs $(\iota_1, \iota_2^T)$, where $\iota_1 : V \to V^*$ and $\iota_2^T : V^* \to V$, so $(\alpha, \mathbf{X})$ will go to $(\iota_1(\mathbf{X}), \iota_2^T(\alpha))$, which will make the equations look like:

(5.3) $$\mathbf{X}' = \iota_2^T(\alpha), \qquad \alpha' = \iota_1(\mathbf{X}).$$

As mentioned above, one way of defining a duality transformation is by way of a scalar product, such as one has with Minkowski space, but not all duality transformations can be put into that form. More generally, one is defining a correlation.

However, if one chooses a frame $\mathbf{e}_\mu$ for $V$ and its reciprocal coframe $\theta^\mu$ then that will define a duality transformation $\iota_\mathbf{e} : V \to V^*$ that amounts to the transposition of the column vector of components of a vector in $V$ with respect to $\mathbf{e}_\mu$ to produce a row vector of components for a covector in $V^*$ with respect to $\theta^\mu$. Any other duality isomorphism can then be obtained by composing that isomorphism with a linear isomorphism of $V^*$:

(5.4) $$\iota = L^T \cdot \iota_{\mathbf{e}} \,.$$

That is because any other duality isomorphism $\iota$ will take the chosen frame $\mathbf{e}_\mu$ to:

(5.5) $$\iota(\mathbf{e}_\mu) = \iota_\nu^\mu \, \theta^\nu \,,$$

in which the matrix $\iota_\nu^\mu$ is invertible. Hence, there as many duality transformations as elements of $GL(n)$. Of course, a duality transformation does not itself belong $GL(n)$.

The group that we shall first consider is the subgroup of $GL(n) \times GL(n)$ that consists of all elements $(L_1, L_2)$ that preserve the canonical bilinear pairing, so:

(5.6) $$<L_2^T(\alpha), L_1(\mathbf{X})> = <\alpha, \mathbf{X}> \qquad \text{for all } \mathbf{X} \in V,\ \alpha \in V^* .$$

When expressed in terms of matrices, that will take the form:

(5.7) $$= \alpha\, L_2\, L_1\, \mathbf{X} = \alpha\, \mathbf{X} \qquad \text{for all } \mathbf{X} \in V,\ \alpha \in V^*,$$

which will imply that:

(5.8) $$L_2\, L_1 = I\,; \qquad \text{i.e.,} \qquad L_2 = L_1^{-1} \,.$$

Thus, the only pairs $(L_1, L_2)$ that preserve the canonical bilinear pairing will have the form $(L_1, L_1^{-1})$, which are then in one-to-one correspondence with the elements of $GL(n)$. One can then think of $GL(n)$ as acting linearly on $V^* \times V$ by taking $(\alpha, \mathbf{X})$ to $(\alpha L^{-1}, L(\mathbf{X}))$.

It is useful to note that since $L_1$ and $L_2$ are both invertible, as long as they both belong to the same conjugacy class, one can always find a unique invertible matrix $A$ that makes $L_2$ take the form:

(5.9) $$L_2 = A\, L_1\, A^{-1} \,.$$

The conjugacy class of $L_1$ is then the orbit of $L_1$ in $GL(n)$ as $A$ ranges over all $A \in GL(n)$. Its isotropy subgroup consists of all $A$ that fix $L_1$ under the action that is defined in (5.9). Hence, all $A$ such that:

(5.10) $$A\, L_1 = L_1\, A.$$

That isotropy subgroup then consists of all invertible matrices that commute with $L_1$, which include not only the scalar multiples of the identity matrix, but also $L_1$ itself, as well as its inverse. Not all invertible matrices are conjugate to each other; in particular, conjugation will preserve eigenvalues, since the eigenvalues of $L_1$ are solutions to the characteristic equation:

$$\det[A L_1 A^{-1} - \lambda I] = \det[A (L_1 - \lambda I) A^{-1}] = \det[L_1 - \lambda I],$$

in which we have factored $I$ into $AA^{-1}$ and used the multiplication rule for determinants. Thus, there will be more than one orbit of $GL(n)$ when it acts upon itself by conjugation; i.e., more than one conjugacy class.

When $L_1$ and $L_2$ do not belong to the same conjugacy class, one can still represent their relationship in the form:

(5.11) $$L_2 = A L_1$$

for some unique invertible $A$ ($= L_2^{-1} L_1$). That will then give one a way of going from one conjugacy class to another.

When $L_2$ is expressed as in (5.9), one can convert the expressions in (5.8) into the form:

(5.12) $$A^{-T} L_1^T A^T L_1 = I; \quad \text{i.e.,} \quad A^{-T} L_1^T A^T = L_1^{-1}.$$

One can also say that:

(5.13) $$L_1^T A^T L_1 = A^T.$$

In this form, one sees that the transformations in question include all of the orthogonal transformations on $V$ when the matrix $A$ defines a scalar product. For instance, when the matrix $A$ is the identity matrix, one will have the Euclidian orthogonal group, and when $A = \eta$ (suitably generalized to dimension $n$), one will have Minkowski space. However, as we mentioned before, that would limit one to only the symmetric invertible matrices, whereas one can now use invertible matrices with more general properties. In particular, when $n$ is even, $A$ can also be antisymmetric, and the linear transformations that it would then define are the symplectic transformations. Hence, we have effectively expanded our scope from scalar products to correlations.

On the other hand, when $A$ is not given any special properties, it can be any element of $GL(n)$, just as $L_1$ can. Hence, the system of equations (5.13) amounts to $n^2$ equations in $n^2$ unknowns, namely, $L_1$. It must then have a unique solution, namely, $I$. Thus, it is only when one looks at more specific correlations, such as ones for which $A$ is symmetric or anti-symmetric, that one will get a non-trivial group that preserves $A$ under the specified action of $GL(n)$.

One then sees that for each choice of $A$, if $L_1$ and $L_2$ satisfy (5.13) then so will the product $L_1 L_2$:

$$L_2^T L_1^T A^T L_1 L_2 = L_2^T A^T L_2 = A^T.$$

Since the identity transformation clearly takes $A^T$ to $A^T$, each choice of $A$ will define a subgroup of $GL(n)$.

As for the duality transformations, one must have:

(5.14) $\qquad < \iota_1(\mathbf{X}), \iota_2^T(\alpha) > = < \alpha, \mathbf{X} > \quad$ for all $\mathbf{X} \in V$, $\alpha \in V^*$,

and in terms of matrices that is:

(5.15) $\qquad (\alpha\, \iota_2\, \iota_1\, \mathbf{X})^T = \alpha\, \iota_2\, \iota_1\, \mathbf{X} = \alpha\, \mathbf{X}\,; \qquad$ i.e., $\qquad \iota_2 = \iota_1^{-1}.$

(The reason that transposing the term in parentheses does not change its value is that it is a scalar.) Thus, one is dealing with essentially the same transformations as in (5.13).

Similarly, any duality transformation $\iota_2$ can be factored into:

(5.16) $\qquad \iota_2 = A^T \iota_1 A^{-1}.$

When $\iota_2$ is expressed in the form (5.16), this will imply that:

(5.17) $\qquad A^{-1} \iota_1^T A\, \iota_1 = I\,; \qquad$ i.e., $\qquad \iota_1^T A\, \iota_1 = A\,.$

The fundamental quadric in $V^* \times V$, for which:

(5.18) $\qquad < \alpha, \mathbf{X} > = 0\,,$

allows one to expand the expand the group of linear transformations that preserve the canonical bilinear pairing to include all pairs $(\lambda, \mu)$ of non-zero scalars that multiply $\alpha$ and $\mathbf{v}$, since:

(5.19) $\qquad < \lambda \alpha, \mu\, \mathbf{X} > = \lambda \mu < \alpha, \mathbf{X} >\,.$

Hence, if $< \alpha, \mathbf{X} >$ vanishes then so will $< \lambda \alpha, \mu\, \mathbf{X} >$. Such transformations are then pairs of homotheties of $V$ and $V^*$ that are centered at their origins.

One can also say that the transformations that preserve the fundamental quadric consist of pairs of equivalence classes $([L], [L^{-1}])$, in which $[L]$ refers to all non-zero scalar multiples of the matrix $L$, which then defines an element of the group $PGL(n)$, which is then the $n$-dimensional *projective group*. It is the group of projective transformations of the projective space $\mathbb{R}P^{n-1}$ that $\mathbb{R}^n - 0$ projects onto by the map that takes every non-zero vector $\mathbf{v}$ in $\mathbb{R}^n$ to the line through the origin $[\mathbf{v}]$ that it generates. Such transformations are also called *homographies*, and they are often written in the form of a system of linear equations that look like:

$$(5.20) \qquad \lambda \bar{X}^\mu = L^\mu_\nu X^\nu,$$

in which $L^\mu_\nu$ is an invertible $n \times n$ matrix, while the components $X^\mu$ and $\bar{X}^\mu$ represent "homogeneous" coordinates of a point in $\mathbb{R}P^{n-1}$.

**6. Discussion.** – One of the obvious extensions of the concepts that were discussed above is to general relativity; that is, to regard the vector spaces in question as tangent and cotangent spaces to a differentiable manifold. In particular, the concept of time+space splittings of the tangent and cotangent bundles is also related to the concept of "static" space-times, which is not surprising since the concept of something being static is closely related to the concept of it being at rest.

When one adopts the viewpoint of pre-metric electromagnetism [**7, 8**], which shifts the center of attention in space-time from the Lorentzian (i.e., metric) structure, which is actually implied by the way that electromagnetic waves propagate in space-time, to the electromagnetic constitutive laws, it also becomes natural to generalize the concept of the Lorentzian structure to other dispersion laws for electromagnetic waves than the traditional classical vacuum law. That would mostly affect the reduction of all transformations of observers and the observer quadric to ones that preserve the dispersion law. However, one should be aware that the higher-degree dispersion laws (such as the quartic one that defines the Fresnel wave surface) will typically have *less* symmetries than the ones that define spatial spheres, just as deforming a sphere to an ellipsoid will reduce its symmetry group from all spatial rotations to only rotations about its axis of rotation as a surface of revolution.

__________